\documentclass[pra,aps,showpacs,floatfix,preprint]{revtex4-1}
\usepackage{color}
\usepackage[utf8]{inputenc}

\usepackage{colordvi}

\usepackage{graphicx,amssymb,epsfig,amsmath}
\usepackage{epstopdf}

\usepackage{amsmath}
\usepackage{bm}
\usepackage{amsfonts}

\usepackage[top=2cm,bottom=2cm,right=2cm,left=2cm]{geometry}

\begin{document}
\title{The multiconfigurational time-dependent Hartree method for bosons with internal degrees of freedom: Theory and composite fragmentation of multi-component Bose-Einstein condensates}

\author{Axel U. J. Lode}
\email{axel.lode@unibas.ch} 
\affiliation{Department of Physics, University of Basel, Klingelbergstrasse 82, CH-4056 Basel, Switzerland}

\begin{abstract}
In this paper the multiconfigurational time-dependent Hartree for bosons method (MCTDHB) is derived for the case of $N$ identical bosons with internal degrees of freedom. The theory for bosons with internal degrees of freedom constitutes a generalization of the MCTDHB method that substantially enriches the many-body physics that can be described. We demonstrate that the numerically exact solution of the time-dependent many-body Schrödinger equation for interacting bosonic particles with internal degrees of freedom is now feasible. We report on the MCTDHB equations of motion for bosons with internal degrees of freedom and their implementation for a general many-body Hamiltonian with one-body and two-body terms that, both, may depend on the internal states of the considered particles. To demonstrate the capabilities of the theory and its software implementation integrated in the MCTDH-X software, we apply MCTDHB to the emergence of fragmentation of parabolically trapped bosons with two internal states: we 
study the groundstate of $N=100$ parabolically confined bosons as a function of the separation between the state-dependent minima of the two parabolic potentials. To quantify the coherence of the system we compute its normalized one-body correlation function. We find that the coherence within each internal state of the atoms is maintained, while it is lost between the different internal states. This is a hallmark of a new kind of fragmentation which is absent in bosons without internal structure. We term the emergent phenomenon ``composite fragmentation''.
\end{abstract}

\maketitle

\section{Introduction}
Since their first creation in ultracold atomic samples of Rubidium \cite{BEC:Rubidium}, Sodium \cite{BEC:Sodium}, and Lithium \cite{BEC:Lithium} in 1995, Bose-Einstein condensates have served scientists as a versatile toolbox for quantum simulation of other quantum systems in, for instance, condensed matter \cite{gauge-exp,top-exp,Greiner,fer2}.
Recent experimental developments with ultracold atoms include the implementation of state-dependent one-body potentials \cite{philipp,gadway,hofferberth}, artificial spin-orbit coupling \cite{spielmanSOC} and gauge fields \cite{gauge-exp,top-exp}. These developments make it necessary to scrutinize several internal degrees of freedom or the hyperfine states of the atoms.   

On the theoretical side, the description of such systems necessitates dealing with the time-dependent many-body Schrödinger equation of interacting, indistinguishable bosonic particles with an internal structure. Popular approaches to the problem in continuous space are the mean-field Gross-Pitaevskii approach \cite{Pethick,Bogoliubov,PitaSandro} and its stochastic variations \cite{TW1,TW2}. For atoms in discretized space or lattices, matrix product states \cite{MPS1,MPS2,MPS3} and the discrete nonlinear Schrödinger equation \cite{PitaSandro} are widespread methods. 

There is to date, however, no ``general'' and in principle exact theoretical description for bosonic atoms with internal structure in continuous space. Here, ``general'' stands for giving reliable predictions for the many-body physics and correlations of interacting bosons with internal structure for any spatial dimension, any interparticle interaction, and any one-body potential.
Most realizations of ultracold bosons with internal degrees of freedom are therefore described with a multi-component Gross-Pitaevskii equation \cite{Matteo,MF1,MF2,MF3,MF4,pete2009}.
In the case of bosonic particles without structure a method which self-consistently incorporates correlations has been devised: the (multi-layer) multiconfigurational time-dependent Hartree for bosons method (MCTDHB) \cite{alon,MCTDHX} (\cite{pete}). In particular, MCTDHB is capable of describing fragmentation \cite{Penrose,Spekkens,Bader,Split} and correlations \cite{RJG,RDMs}, i.e., samples of ultracold bosons where the reduced one-body density matrix has several eigenvalues of the order of the particle number \cite{RDMs,Penrose}. Ref.~\cite{MCTDHX} mentions the applicability of the MCTDHB to particles with internal degrees of freedom, but does not further discussed or show an implementation of it. Moreover, MCTDHB is numerically exact: it can solve the time-dependent many-body Schrödinger equation (TDSE) to an arbitrary large degree of precision \cite{Axel_exact,Axel_book}. Importantly, fragmentation and thereby correlations which are neglected in mean-field approaches like the Gross-Pitaevskii equation, 
are known to be present in spinor Bose-Einstein condensates (BECs) \cite{mueller,sun:dynfrag,song:SOCfrag,Ho:frag}. For instance, it has been shown that fragmentation may 
emerge dynamically \cite{sun:dynfrag}, and that it is present in systems with spin-orbit coupling \cite{song:SOCfrag} or systems with angular momentum \cite{mueller,Ho:frag}. This renders an in principle exact method like MCTDHB which can make accurate predictions for the many-body correlations of fragmented systems with internal degrees of freedom like spin or a level-structure a much needed and versatile tool.

The present paper derives the MCTDHB equations of motion for the case of bosons with an internal structure which may be seen as a generalization of the multi-orbital mean-field theory developed in Ref.~\cite{MOMF} to the realm of many-body physics.
The capabilities of the derived theory are demonstrated by exhibiting the emergence of fragmentation and correlations in a system with one-body potentials which are dependent on the internal state of the considered particles similar to the experimentally realized one in Refs.~\cite{philipp,Matteo}. A new kind of fragmentation which is qualitatively different from the single-component fragmentation of bosonic systems without internal degrees of freedom is found: composite fragmentation -- the system's components maintain their coherence while the coherence between them is lost.

We mention here, that the software implementation of the theory developed in this paper has been incorporated into the MCTDH-X software and is freely available \cite{ultracold}.

The structure of the paper is as follows: in Section \ref{MCTDHB}, the equations of motion of MCTDHB for particles with internal degrees of freedom are derived, in Section \ref{deltafrag} the emergence of fragmentation and correlations in the groundstate of a system with state-dependent one-body potentials is demonstrated. Conclusions and an outlook are given in Section \ref{conc}.

\section{MCTDHB for bosons with internal degrees of freedom}\label{MCTDHB}
\subsection{Schrödinger equation and Hamiltonian}
Our task is the solution of the time-dependent many-boson Schrödinger equation 
\begin{equation}
 \hat{H} \vert \Psi \rangle = i \partial_t \vert \Psi \rangle \label{TDSE}
\end{equation}
for a state $\vert \Psi \rangle$ of interacting indistinguishable bosons with internal structure.
It is assumed that the many-body Hamiltonian $\hat{H}$ contains one-body and two-body operators, $\hat{\underline{h}}$ and $\hat{\underline{W}}$, respectively. There is a one-body term for each boson and a two-body term for every pair of bosons, i.e.,
\begin{equation}
 \hat{H}= \sum_{i=1}^N \hat{\underline{h}} (\vec{r}_i;t) + \sum_{i<j=1}^N \hat{\underline{W}}(\vec{r}_i, \vec{r}_j;t). \label{H1}
\end{equation}
The $\underline{\cdot}$ notation indicates the vector or spinor character of the operators. The one-body Hamiltonian $\hat{\underline{h}}$ as well as the two-body Hamiltonian $\hat{\underline{W}}$ can be different for each internal degree of freedom and may contain couplings between them. For particles with $\xi$ internal degrees of freedom one may write
\begin{equation}
 \hat{H} = \sum_{i=1}^N \sum_{\alpha,\alpha'=1}^\xi \hat{h}_{\alpha,\alpha'} (\vec{r}_i;t)  \mathbf{1}^\alpha_i \mathbf{1}^{\alpha',T}_i + \sum_{i<j=1}^N \sum_{\stackrel{\alpha,\alpha',}{\alpha'',\alpha'''=1}}^\xi \hat{W}_{\alpha,\alpha',\alpha'',\alpha'''}(\vec{r}_i, \vec{r}_j;t) \mathbf{1}^\alpha_{i}\mathbf{1}^{\alpha',T}_{i}\mathbf{1}^{\alpha''}_{j}\mathbf{1}^{\alpha''',T}_{j}. \label{Honehalf}
\end{equation}
Here, $\mathbf{1}^\alpha_k$ ($\mathbf{1}^{\alpha,T}_k$) is the (transpose of the) $\alpha$-th unit vector in the space of the internal degrees of freedom for the $k$-th particle. This ensures, that the respective operators acts on the $\alpha$-th degree of freedom. Here and in the following, the index $\alpha$ is used for the internal degrees of freedom of the considered atoms. With the notation of Eq.~\eqref{Honehalf} it becomes clear that the one-body term $\hat{h}_{\alpha,\alpha'}$ transfers particles from internal state $\alpha$ to $\alpha'$ and that the two-body term 
$\hat{W}_{\alpha,\alpha',\alpha'',\alpha'''}$ transfers particles from internal state $\alpha$ to $\alpha'$ and from $\alpha''$ to $\alpha'''$.
 In second quantized notation, the state $\vert \Psi \rangle$ corresponds to a field operator,
\begin{equation}
 \hat{\Psi}^\dagger(\vec{r},t) = \sum_k \hat{b}^\dagger_k  \vec{\varphi}^*_k(\vec{r};t) \equiv \sum_k \hat{b}^\dagger_k \left( \sum_{\alpha=1}^\xi \phi^{\alpha,*}_k (\vec{r};t)  \mathbf{1}^\alpha \right).
\end{equation}  
Here, and in the following we employ the vector notation $\vec{\varphi}$ for multi-level orbitals and the symbol $\phi$ for their components.
The orbitals with internal degrees of freedom $\vec{\varphi}_k(\vec{r};t)$ form an orthonormal and time-dependent basis.
When we express the Hamiltonian in Eq.~(\ref{H1}) with the operators $\lbrace \hat{b}_k,\hat{b}^\dagger_k\rbrace$ it takes on the form,
\begin{equation}
 \hat{H}= \sum_{k,q} h_{kq} \hat{b}^\dagger_k  \hat{b}_q + \sum_{kqsl} W_{ksql} \hat{b}^\dagger_k \hat{b}^\dagger_s \hat{b}_q \hat{b}_l. \label{H2}
\end{equation}
It is worthwhile to note here, that this Hamiltonian is of the same form as in the case of structureless bosons, and hence also the derivation of the MCTDHB equations of motion will be similar to the case of structureless bosons (see following subsection).
Here, $h_{kq}$ are the matrix elements of the one-body Hamiltonian $\hat{\underline{h}}$,
\begin{equation}
 h_{kq} = \langle \vec{\varphi}_k \vert \hat{\underline{h}} \vert \vec{\varphi}_q \rangle  = \sum_{\alpha,\alpha'} \langle \phi^\alpha_k \vert \hat{h}_{\alpha,\alpha'} \vert \phi^{\alpha'}_q \rangle,  \label{onebody-elements}
\end{equation}
 and $W_{ksql}$ are the matrix elements of the two-body operator $\hat{\underline{W}}$,
\begin{eqnarray}
 W_{ksql}&=& \iint d\vec{r} d\vec{r}'  \vec{\varphi}^*_k (\vec{r};t) \vec{\varphi}^*_s (\vec{r}';t)  \hat{\underline{W}}(\vec{r},\vec{r}';t)  \vec{\varphi}_q (\vec{r};t) \vec{\varphi}_l (\vec{r}';t). \label{wksql}
\end{eqnarray}
The final prerequisite for the derivation of the MCTDHB equations of motion are the matrix elements of the reduced one-body and two-body density matrices,
\begin{eqnarray}
\rho_{kq} &=& \langle \Psi \vert \hat{b}^\dagger_k \hat{b}_q \vert \Psi \rangle, \label{R1E} \\ 
\rho_{ksql} &=& \langle \Psi \vert \hat{b}^\dagger_k \hat{b}^\dagger_s \hat{b}_q \hat{b}_l \vert \Psi \rangle. \label{R2E}         
\end{eqnarray}
Importantly, all the quantities [Eqs.~\eqref{H2},\eqref{onebody-elements},\eqref{wksql},\eqref{R1E},\eqref{R2E}] that are of relevance to the derivation of the MCTDHB equations of motion (cf. Ref.~\cite{alon}) can be cast in a form that is identical to the case of bosons without internal structure.

\subsection{Derivation of the MCTDHB equations of motion}
The MCTDHB method uses a multiconfigurational time-adaptive ansatz and many-body basis to tackle the TDSE \eqref{TDSE},
\begin{eqnarray}
 \vert \Psi \rangle &= \sum_{\vec{n}} C_{\vec{n}} (t) \vert \vec{n} ; t \rangle, \label{ansatz} \\
 \vert \vec{n} ; t \rangle &= \prod_{i=1}^M \left[ \frac{\left( \hat{b}^\dagger_i(t) \right)^{n_i}}{ \sqrt{n_i!}} \right] \vert vac \rangle \label{basis}.  
\end{eqnarray}
Here, $\vec{n}$ is an occupation number vector, $\vec{n}=(n_1, ..., n_M)$, of $M$ time-dependent and mutually orthonormal multilevel orbitals or spinors $\vec{\varphi}_k(\vec{r};t)$ that are created by the operators $\hat{b}^\dagger_k(t)$,

For spin-$\frac{1}{2}$ bosons with two internal degrees of freedom, 
for instance, the orbitals $\vec{\varphi}_k$ take on the following form:
\begin{equation}
\vec{\varphi}_k(\vec{r};t) = \sum_{\alpha=+,-} \mathbf{1}^{\alpha} \phi_k^{\alpha}(\vec{r};t) = \left(
 \begin{array}{c}
  \phi^+_k(\vec{r};t) \\
  \phi^-_k(\vec{r};t) \\
 \end{array}
\right).
\end{equation}
Note, that there is no relative phase between the different components $\phi_k^\alpha$ of the orbital $\vec{\varphi}_k$: the components are described by general and unrelated time-dependent functions $\phi_k^\alpha(\vec{r};t)$. As we will see in the following, the components of different orbitals are coupled through the orthonormality constraint.
The functional action of the TDSE, including Lagrange multipliers $\mu_{ij}(t)$ to ensure the orthonormalization of the time-dependent multilevel orbitals, reads
 \begin{equation}
\mathcal{S} = \int dt \left( \langle \Psi \vert \hat{H}- i\partial_t \vert 
\Psi \rangle+ \sum_{ij} \mu_{ij}(t)\left( \langle \vec{\varphi_i} \vert  \vec{\varphi_j} 
\rangle - \delta_{ij} \right) \right). \label{action} 
\end{equation}
Since the action functional is of the same shape as the one in the case of bosons without internal structure, the equations of motion are also of identical shape. They are stated here, for the sake of completeness and using the invariance property  $\langle  \vec{\varphi}_j \vert \partial_t \vert \vec{\varphi}_k\rangle =0$ (see Ref.~\cite{invariance}) to simplify their appearance. For the details of the derivation see, for instance, Ref.~\cite{alon}.
\begin{eqnarray}
 i \partial_t \mathcal{C} &=& \mathcal{H} \mathcal{C}; \qquad \mathcal{H}_{\vec{n}\vec{n}'} = \langle \vec{n}' ; t \vert \hat{H} \vert \vec{n} ; t \rangle, \label{CEOM} \\
 i \partial_t \vert \vec{\varphi}_j \rangle &=& \mathbf{P} \left[ \hat{\underline{h}} \vert \vec{\varphi}_j \rangle + \sum_{ksql} \lbrace \rho_{jk} \rbrace^{(-1)} \rho_{ksql} \hat{\underline{W}}_{sl}(\vec{r},t) \vert \vec{\varphi}_q  \rangle \right], \label{OEOM} \\
 \mathbf{P} &=& \underline{\mathbf{1}} - \sum_k  \vert \vec{\varphi}_k \rangle \langle \vec{\varphi}_k \vert .
\end{eqnarray}
Here, the local interaction potentials 
\begin{equation}
 \hat{\underline{W}}_{sl}(\vec{r},t) = \int d\vec{r}' \vec{\varphi}^*_s(\vec{r}',t) \hat{\underline{W}}(\vec{r},\vec{r}',t) \vec{\varphi}_l(\vec{r}',t)
\end{equation}
were defined. 
The above equations \eqref{CEOM} and \eqref{OEOM} form the heart of MCTDHB for bosons with internal structure. The $M$ multilevel orbitals' equations are a set of coupled, nonlinear and integro-differential equations of motion. The $N_{\text{conf}}=\binom{N+M-1}{N}$ equations of motion of the coefficients are linear and coupled to the orbitals' equations since the application of the Hamiltonian necessitates the matrix elements $h_{kq}(t)$ and $W_{ksql}(t)$ that are in turn functions of the orbitals $\vec{\varphi}_k$. The orbitals equations are coupled to the coefficients' ones because the matrix elements of the reduced one- and two-body densities, $\rho_{kq}(t)$ and $\rho_{ksql}(t)$, are functions of the coefficients.
An important distinction in the equations of motion from the case of structureless bosons \cite{alon} emerges in the projection operator: the different orthonormality relations for multileveled orbitals result in a more involved projection operator. One may rewrite $\mathbf{P}$ as follows for the case of $\xi$-leveled particles
\begin{equation}
 \mathbf{P} = \underline{\mathbf{1}} - \Bigg[ \sum_{k=1}^M \sum_{\alpha,\alpha'=1}^\xi  \mathbf{1}^\alpha \vert \phi^\alpha_k \rangle \langle \phi^{\alpha'}_k \vert \mathbf{1}^{\alpha',T} \Bigg].
\end{equation}
In this notation, it becomes clear that the projection operator is built up from different and orthonormal spinor orbitals.
Here, $\underline{\mathbf{1}}$ denotes the unit matrix in the internal degrees of freedom.
Let us comment here on the solution of the equations of motion \eqref{CEOM} and \eqref{OEOM} for time-evolutions and eigenstates. Consider a Hamiltonian that preserves the total spin or the number of bosons in the different components of the system. The time-evolution generated by such a Hamiltonian, will of course also preserve the number of atoms in each component. For the computation of the ground states investigated in the following Section, Eqs.~\eqref{CEOM} and \eqref{OEOM} are propagated in imaginary time $t\rightarrow -i\tau$, such that unwanted excitations are damped out exponentially. After every imaginary time propagation step, the wavefunction is re-normalized. The excitations may have different numbers of particles in their components as the spectrum of the Hamiltonian generally contains all possible distributions of particles among components. The number of particles in each component is hence not conserved in the process of imaginary time propagation. In this way, it is guaranteed that the 
wavefunction with a distribution of particles between the components that minimizes the total energy of the system is obtained as result of the imaginary time propagation.
This concludes the exhibition of the MCTDHB equations of motion for bosons with internal structure. We move on to an application of the derived method to bosons in state-dependent one-body potentials.

\section{Emergence of fragmentation in spin $\frac{1}{2}$ bosons in state-dependent potentials}\label{deltafrag}
The experiment in Ref.~\cite{philipp,Matteo} realizes a Bose-Einstein condensate of atoms with two internal degrees of freedom which is confined in three-dimensional parabolic traps the minima of which have different positions for the two internal states of the atoms. In Ref.~\cite{philipp} the atoms are dynamically and reversibly entangled using a separation and recombination scheme and in Ref.~\cite{Matteo} the excitation spectrum of the system is dynamically probed using sideband Rabi spectroscopy demonstrating a temperature of less than $30$nK.
Motivated by these experiments, we apply MCTDHB to a system of $N=100$ one-dimensional bosons that have spin $\frac{1}{2}$ and are governed by a Hamiltonian which contains both spin-dependent and spin-independent interparticle interactions. We compute the ground-state densities, fragmentation, and one-body correlation functions of the system as a function of the minima of the state-dependent potentials. 

\subsection{System Hamiltonian}
\begin{figure}
\includegraphics[width=\textwidth]{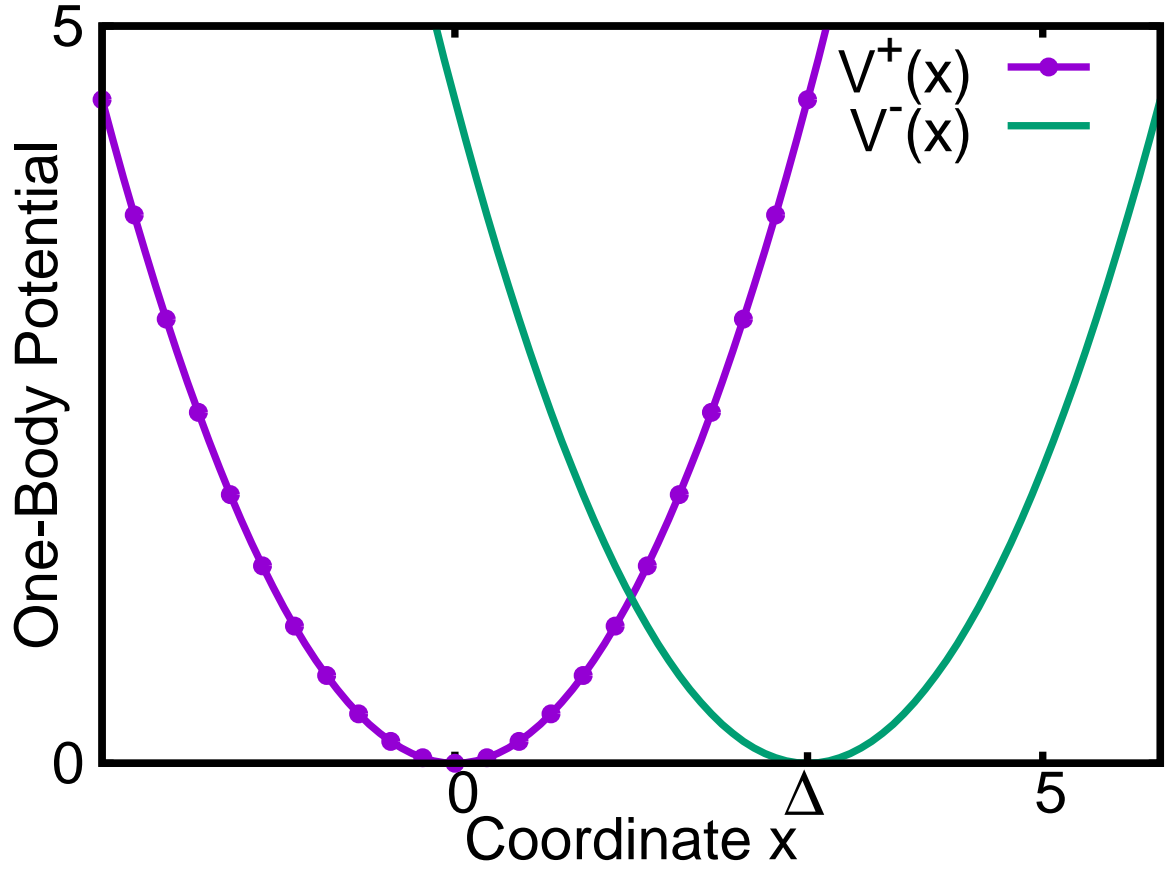} 
\caption{State-dependent potentials. In each internal state ($\alpha=+$ and $\alpha=-$) of the atoms the potential is harmonic. The minima of the potentials are displaced by the separation parameter $\Delta$. All quantities shown are dimensionless.}
\label{Fig:Pot}
\end{figure}

We start our investigation by specifying the Hamiltonian of the spinor bosons that we consider in this section.
In the following, we consider two internal states and label them $\alpha=+$ and $\alpha=-$, respectively. Since we consider a quasi one-dimensional system, we will furthermore use the label $x$ instead of $\vec{r}$ for coordinates.
For the sake of simplicity, we assume that the spatial dependence of the inter-particle interaction is the same time-independent contact potential for all components or levels, i.e., $\hat{W} (\vec{x},\vec{x}') = \delta(\vec{r}-\vec{r}') \; \forall \alpha$. Furthermore, we employ dimensionless units for the sake of computational convenience. This means, we divide the Hamiltonian by $\hbar^2 (m L^2)^{-1}$, where $m$ is the mass of the considered particles and $L$ is a conveniently chosen length-scale. Explicitly, we find
\begin{eqnarray}
 \hat{\underline{h}}(x;t) &=& \left[ -\frac{1}{2} \underline{\partial}^2_{x} + \underline{V}(x,t) \right] \label{one-body} \\
 &=& \sum_{\alpha=+,-} \left[ -\frac{1}{2} \partial^2_{x} +  V_\alpha (x,t) \right]  \mathbf{1}^\alpha \nonumber \\
 \hat{\underline{W}}(x,x') &=& \sum_{\alpha=+,-} \left( \mathbf{1}^\alpha \mathbf{1}^{\alpha,T}  \lambda^\alpha_0 \delta(x-x') \right) + \lambda_1 \left( \sum_{\nu = 1,2,3} \mathbf{S}^{(x)}_\nu \otimes  \mathbf{S}^{(x')}_\nu
 \right)   \delta(x-x') . \label{H3}
\end{eqnarray}
As introduced previously, the $\mathbf{1}^{\alpha}$ ($\mathbf{1}^{\alpha,T}$) is the (transpose of the) $\alpha$-th unit vector in the space of the internal degrees of freedom that takes care that the respective operators act on each component of the spinors, respectively. 
The one-body potentials $V_\alpha(\vec{r})$ are given by
\begin{equation}
V_-(x)=\frac{1}{2} x^2; \qquad V_+(x)=\frac{1}{2} (x-\Delta)^2\label{deltapot}
\end{equation}
Here, $\Delta$ is the distance or separation of the minima of the parabolic potentials of the $+$ and $-$ internal states of the atoms. See Fig.~\ref{Fig:Pot} for a plot of the above potentials. 

The interparticle interaction in Eq.~\eqref{H3} contains a spin-independent contribution with the interaction strength $\lambda^\alpha_0$ that acts identically on each of the spinor components. Additionally, a spin-dependent contribution is present in $\hat{\underline{W}}$. It is scaled with the interaction strength $\lambda_1$. The $\mathbf{S}^{(x)}_{1/2/3}$ are a realization of a spin algebra that mediate the spin-dependent interparticle interaction. Where present, the superscript indicates the coordinate space in which the spin operator acts.

The local interaction potentials [cf. Eq.~(17)] for the interaction in the Hamiltonian in Eq.~\eqref{H3} read explicitly:
\begin{equation}
 \hat{\underline{W}}_{sl}(\vec{r},t) = \sum_{\alpha=+,-} \lambda_0^\alpha \phi^{*,\alpha}_s(x,t)\phi^\alpha_l(x,t)
 + \lambda_1 \sum_{\nu=1,2,3} \left[ \vec{\varphi}^*_s(x,t)  \mathbf{S}^{(x)}_\nu \vec{\varphi}_l(x,t)  \right] \cdot \mathbf{S}^{(x)}_\nu.
\end{equation}

In the absence of interparticle interactions, i.e., if $\lambda_0^+=\lambda_0^-=\lambda_1=0$, the Hamiltonian in Eq.~\eqref{one-body} with the potential in Eq.~\eqref{deltapot} yields the same energy, irrespective of the distribution of particles among the states $+$ and $-$: the ground state is highly degenerate. The scattering behavior of particles in different internal states is typically slightly different, i.e., $\lambda_0^+\neq \lambda_0^-$. This lifts the mentioned degeneracy and renders the ground state of the system to be a condensate in the internal state with the smaller repulsive interaction strength, see also Sec.~\ref{Sec:Results} below. 

Finally, we note that Ref.~\cite{rbscattering} determined the $s$-wave scattering lengths for a three-dimensional two-component Bose-Einstein condensate of $^{87}$Rb from its collective oscillations. In accordance with this Reference ~\cite{rbscattering}, we chose the one-dimensional interaction strengths $\lambda^+_0 = 0.01; \lambda^-_0=0.00975; \lambda_1=0.0095$. For an example of parameters to achieve these quasi-one-dimensional interaction strengths from the three-dimensional study in Ref.~\cite{rbscattering}, see Ref.~\cite{units}.

\subsection{Quantities of Interest}
The observables that we are going to use to analyze the ground states of $N=100$ bosons in the state dependent trapping potentials of Fig.~\ref{Fig:Pot} are the one-body density, the fragmentation, and the normalized one-body correlation functions. Here and in the following, we will omit the dependence of quantities on time for notational convenience and because we are going to investigate eigenstates of a system and not its dynamics. We will furthermore use bold math symbols for quantities that have the internal degrees of freedom of the system and standard math symbols for quantities which do not have internal degrees of freedom.

The one-body density $\bm{\rho}(x)$ can be computed from the matrix elements of the reduced one-body density matrix and the orbitals $\vec{\varphi}_k(x)$ as follows
\begin{equation}
 \bm{\rho}(x)=\sum_{kq\alpha} \rho_{kq}  \mathbf{1}^\alpha \phi^{\alpha,*}_k(x) \phi^\alpha_q(x).
\end{equation}
Since the one-body density is a vector of densities in every internal degree of freedom of the considered particles, it is instructive to define the component and composite densities
\begin{equation}
 \rho^\alpha(x)= \mathbf{1}^\alpha \bm{\rho}(x)=\sum_{kq} \rho_{kq} \phi^{\alpha,*}_k(x) \phi^\alpha_q(x)., \label{componenteRho}
\end{equation}
and 
\begin{equation}
\rho^{+/-}(x)= \sum_\alpha \mathbf{1}_x^\alpha \bm{\rho}(x)=\sum_{kq\alpha} \rho_{kq} \phi^{\alpha,*}_k(x) \phi^\alpha_q(x)., \label{compositeRho}
\end{equation}
respectively. Here, $\mathbf{1}_x^\alpha$ indicates that each component $\alpha$ is projected to the same spatial coordinate $x$ to form $\rho^{+/-}(x)$.
The density $\bm{\rho}(x)$ quantifies the probability for all internal states, respectively, to find a particle at position $x$. The composite density $\rho^{+/-}(x)$ defines the probability to find a boson at position $x$ irrespective of its internal state and the component densities $\rho^{\alpha}(x)$ define the probability to find a particle in internal state $\alpha$ at position $x$. 

Similar to the one-body density $\bm{\rho}(x)$, the normalized one-body correlation function $\bm{g}^{(1)}$ is a multi-component quantity. Its definition \cite{RDMs,RJG} is
\begin{equation}
 \bm{g}^{(1)}(x,x')= \frac{\bm{\rho}^{(1)}(x,x')}{\sqrt{\bm{\rho}^{(1)}(x,x)\bm{\rho}^{(1)}(x',x')}}, 
\end{equation}
where 
\begin{equation}
 \bm{\rho}^{(1)}(x,x')=\sum_{kq\alpha} \rho_{kq} \mathbf{1}^\alpha \phi^{\alpha,*}_k(x') \phi^{\alpha,*}_q(x) 
\end{equation}
is the reduced one-body density matrix.
Analogous to the composite and component one-body densities, we define the component and composite normalized one-body correlation functions,
\begin{equation}
 g^{(1),\alpha}(x,x')= \frac{\rho^{(1),\alpha}(x,x')}{\sqrt{\rho^{(1),\alpha}(x,x)\rho^{(1),\alpha}(x',x')}},  
\end{equation}
and
\begin{equation}
 g^{(1),+/-}(x,x')= \frac{\rho^{(1),+/-}(x,x')}{\sqrt{\rho^{(1),+/-}(x,x)\rho^{(1),+/-}(x',x')}},  
\end{equation}
respectively. Here, the component reduced one-body density matrix, $\rho^{(1),\alpha}=\mathbf{1}^\alpha \bm{\rho}^{(1)}$ as well as the composite one-body density matrix, $\rho^{(1),+/-}=\sum_\alpha \mathbf{1}^\alpha_{x,x'} \rho^{(1),\alpha}(x,x')$ , were used. Here, $\mathbf{1}^\alpha_{(x,x')}$ indicates that each component is projected to the same set of spatial coordinates $(x,x')$ to obtain $\rho^{(1),+/-}$.
The normalized one-body correlation function $g^{(1)}$ quantifies the coherence of the bosons and can be measured for instance in interference experiments \cite{corr1,corr2}. Let us note here that all, the total, composite, and component normalized one-body correlation functions are fixed to unity in the case of a mean-field state of the system \cite{RJG,RDMs}. Therefore, the one-body correlation functions may be used to quantify the failure of a mean-field description to describe a given system's state. To summarize, $\bm{g}^{(1)}(x,x')$ gives a spatially resolved picture of how well the reduced density matrix $\bm{\rho}^{(1)}(x,x')$ can be described by a single complex valued function, i.e., a mean-field approach at $x,x'$.

It remains to define fragmentation, which quantifies what fraction of the bosons does not occupy the eigenfunction of the reduced one-body density matrix or natural orbital corresponding to the largest eigenvalue. To determine the fragmentation $F$, we hence determine the eigenvalues of the matrix-elements $\rho_{kq}$ of the reduced one-body density matrix [Eq.~\eqref{R1E}], the so-called natural occupations $\rho^{(NO)}_kq$:
\begin{equation}
 F=\sum^M_{j=2} \rho^{(NO)}_j = 1-\rho^{(NO)}_1 \label{Frag}
\end{equation}
This concludes the exhibition of the quantities of interest and we now move on to the discussion of the results.

\subsection{Results}\label{Sec:Results}
\begin{figure}
\includegraphics[width=\textwidth,angle=-90]{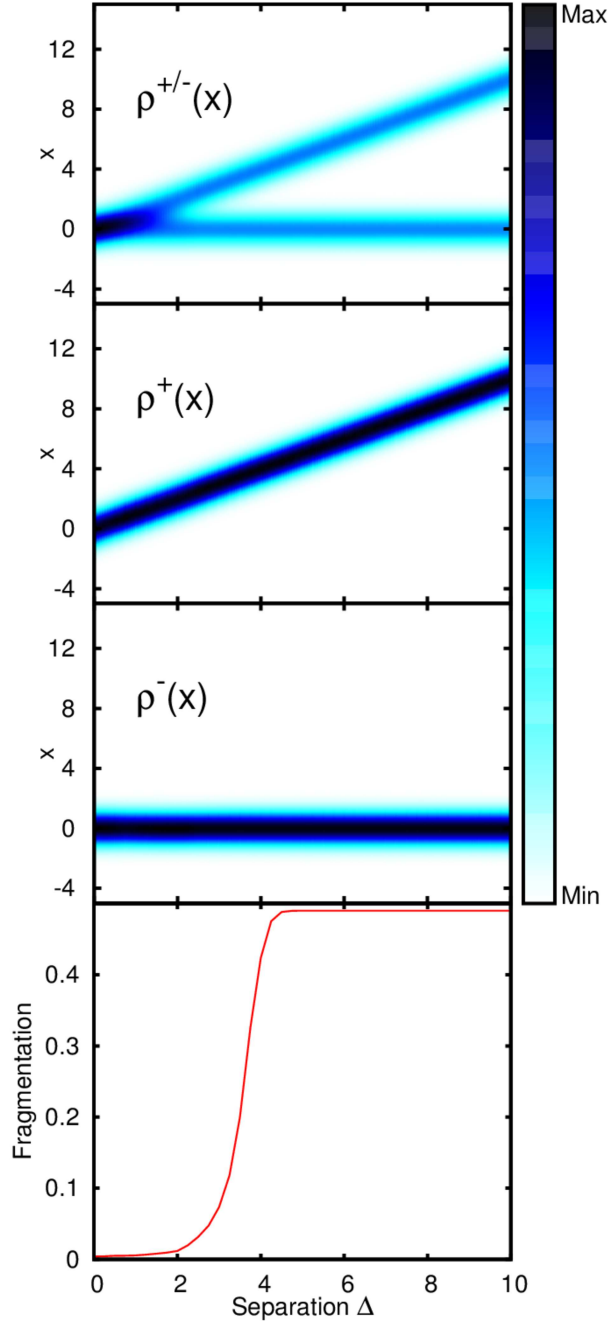}
\caption{Groundstate density and fragmentation as a function of the separation $\Delta$. The top panel shows the composite density $\rho^{+/-}(x)$ of the two internal states $\alpha=+$ and $\alpha=-$. The second and third panel depict the component densities $\rho^{\alpha}(x)$ of the system in the respective internal state $\alpha=+$ and $\alpha=-$. The bottom panel shows the fragmentation of the system. Fragmentation is energetically favorable as soon as the overlap of the densities of the internal states becomes small (cf. top and bottom panels). All quantities shown are dimensionless.}
\label{Fig:Densities}
\end{figure}

We now set out to analyze how the physics of a sample of $N=100$ bosons in the state-dependent potentials $V_\alpha(x)$ depends on the separation $\Delta$. We use $M=3$ orbitals in the present study which yields a problem set including $5005$ coefficients. By testing with $M=4$ and $176851$ coefficients for the fully fragmented cases with large $\Delta$, we assessed the convergence of our results with respect to the number of orbitals. In our simulations, we used a discrete variable representation \cite{review} of $512$ functions on a grid of extent $[-10,20]$. We checked the exactness of our grid representation by making sure that the densities are less than $10^{-10}$ of their maximal values on the edges of the grid. Furthermore, we assessed that the energy differs by a factor less than $10^{-10}$ in computations with $512$ and $1024$ functions in the discrete variable representation.
Since the results below do not change anymore when we increase the number of variational parameters (orbitals $M$) or the number of grid points, they are \textit{numerically exact}. We commence the analysis by first computing the component and composite densities [cf. Eqs.~\eqref{componenteRho},\eqref{compositeRho}] as well as the fragmentation [cf. Eq.~\eqref{Frag}] and plot them in Fig.~\ref{Fig:Densities}. The component and composite densities follow an intuitive pattern: the component densities retain their Gaussian shape irrespective of the separation $\Delta$, but their maxima follow the $\Delta$-dependent minima of the state-dependent potentials. Since the repulsion in state $\alpha=-$ is slightly weaker than in state $\alpha=+$, the number of bosons in state $\alpha=-$ is roughly $51$, i.e., slightly larger than that of state $\alpha=+$ which is roughly $49$ for small separations $\Delta$. For larger separations, this imbalance disappears gradually and is gone from $\Delta\gtrsim4$. Already in this 
simple example, fragmentation emerges once the separation becomes larger than $\Delta\approx 3$ (see lowest panel of Fig.~\ref{Fig:Densities}). This is in stark contrast to bosonic atoms with contact interactions and without internal degrees of freedom, where fragmentation in single well traps is almost absent \cite{Sascha,Budha}. In the present case of atoms with internal degrees of freedom, weak interactions are sufficient to yield a fully two-fold fragmented state for $\Delta \gtrsim 4$, where $F\approx 0.5$ and $\rho^{(NO)}_1\approx\rho^{(NO)}_2\approx 0.5$ while $\rho_k^{(NO)}\approx0\;\forall\;k\geq3$. This behavior of the fragmentation resembles the case of bosons without internal degrees of freedom in a double well when the height of the barrier in the center is increased \cite{RDMs,KasparBJJ,Split}: there are precisely two significant natural occupations and, as mentioned, the others are zero. A marked difference between the single-component system in a double well and the present case of bosons 
with 
internal degrees of freedom which feel different one-body potentials is the absence of a potential barrier: the atoms reside in single wells, but in distinct internal states. Furthermore, there is a minimal imbalance of $\lesssim 1\%$ of the number of atoms in the $\alpha=+$ and the $\alpha=-$ internal state of the atoms, because the scattering rate $\lambda_1^+$ is slightly larger than $\lambda_1^-$. For fragmented, structureless bosons in a symmetric double well, there is no such imbalance. It is of further interest to determine if the two-fold fragmentation of the system is due to the macroscopic occupation of two orbitals $\vec{\varphi}_1,\vec{\varphi}_2$ which have non-vanishing contributions in both their components $\phi_k^+,\phi_k^-$ for $k=1,2$ or if in some orbitals the contributions of one component vanishes, i.e., if $\int \vert \phi_k^\alpha(x) \vert^2 dx \approx 0$ holds for some $k,\alpha$. To asses this, one may investigate correlation functions or plot the orbitals' components. We defer a 
detailed analysis of the coefficients and orbitals which build up the many-body wavefunction to the Appendix and move on to investigate the spatial one-body correlation functions $\bm{g}^{(1)}(x,x')$ of the system encompassing its fragmentation. 

To get a spatially- and state-resolved picture of fragmentation in the two-component system, we plot the composite and component correlation functions $\vert g^{(1),+/-}(x,x')\vert^2$,$\vert g^{(1),+}(x,x')\vert^2$, and $\vert g^{(1),-}(x,x')\vert^2$, respectively, for various separations $\Delta$ in Fig.~\ref{Fig:Corr}. When the separation is close to zero, the coherence of the sample is maintained in all space, since both, the component and composite correlation functions are almost unity, i.e., $\vert g^{(1),+} \vert^2 \approx \vert g^{(1),-} \vert^2 \approx \vert g^{(1),+/-}\vert^2 \approx 1$ (see top row of panels in Fig.~\ref{Fig:Corr}). As soon as the separation $\Delta$ is increased, the off-diagonal of the composite correlation function starts to drop to zero rapidly -- the component correlation functions, however, show clearly that the coherence within the two 
internal states of the atoms is maintained, i.e. $\vert g^{(1),+} \vert^2 \approx \vert g^{(1),-} \vert^2 \approx 1$ (see middle row of Fig.~\ref{Fig:Corr}). When the system is fully split, the components still maintain their coherence while the composite coherence is almost completely gone (see bottom row of Fig.~\ref{Fig:Corr}).  
The fragmentation observed here hence differs qualitatively from fragmentation in the case of bosonic particles without internal structure: the coherence is lost not between the atoms in one component, but between the atoms in distinct components of the system. We hence term this kind of fragmentation ``composite fragmentation'' as opposed to ``component fragmentation'' which can also be seen for single-component systems (see for instance Refs.~\cite{KasparBJJ,Bader,Split}). The composite correlation function bears some resemblance to the case of bosons without internal degrees of freedom in a double well with a large barrier \cite{RDMs}. Since composite fragmentation emerges when the spatial overlap of the component densities becomes small, we infer that this triggers fragmentation in the present case.

\section{Conclusions}\label{conc}
The theory described here for bosons with internal structure constitutes a generalization of the MCTDHB method which substantially enriches the many-body physics that can be described with the approach. Since MCTDHB is a method that is in principle exact -- once convergence with the number of variational parameters is achieved, the result is a solution of the full time-dependent many-body problem -- numerically exact solutions of the time-dependent many-body Schr\"odinger equation for interacting bosons with internal degrees of freedom are enabled by this work. Moreover, the software implementation of MCTDHB for systems with internal degrees of freedom which was used to obtain the results in this work is incorporated in the MCTDH-X software and openly available \cite{ultracold}.

The emergence of fragmentation was found in the ground state of $N=100$ bosons, when the minima of their state-dependent parabolic one-body potentials are taken apart. Interestingly, the emergent fragmentation is visible in a decreased coherence quantified by the composite correlation function and absent in the component correlation functions. Such a buildup of correlations and loss of coherence cannot be present in single-component systems, because these can have correlations only between atoms in distinct orbitals and not between atoms in distinct components. We hence term the emergent phenomenon composite fragmentation as opposed to component fragmentation which may also be present for bosonic particles without internal structure.

As further directions, we would like to mention here the application of MCTDHB to non-equilibrium, i.e., dynamical systems and to further scrutinize and assess the physics of the interplay of composite and component fragmentation of time-independent and time-dependent systems.

\acknowledgments{Fruitful discussions with and useful reference suggestions by Matteo Fadel are acknowledged. Insightful discussions with Ofir E. Alon about the projection operator, comments and discussions about the manuscript with Elke Fasshauer, Marios C. Tsatsos, and Christoph Bruder as well as financial support by the Swiss SNF and the NCCR Quantum Science and Technology and computation time on the Hornet and Hazel Hen clusters of the HLRS in Stuttgart are gratefully acknowledged.}

\appendix*
\section{Detailed analysis of the many-body wavefunction}
\begin{figure}
 \includegraphics[width=\textwidth]{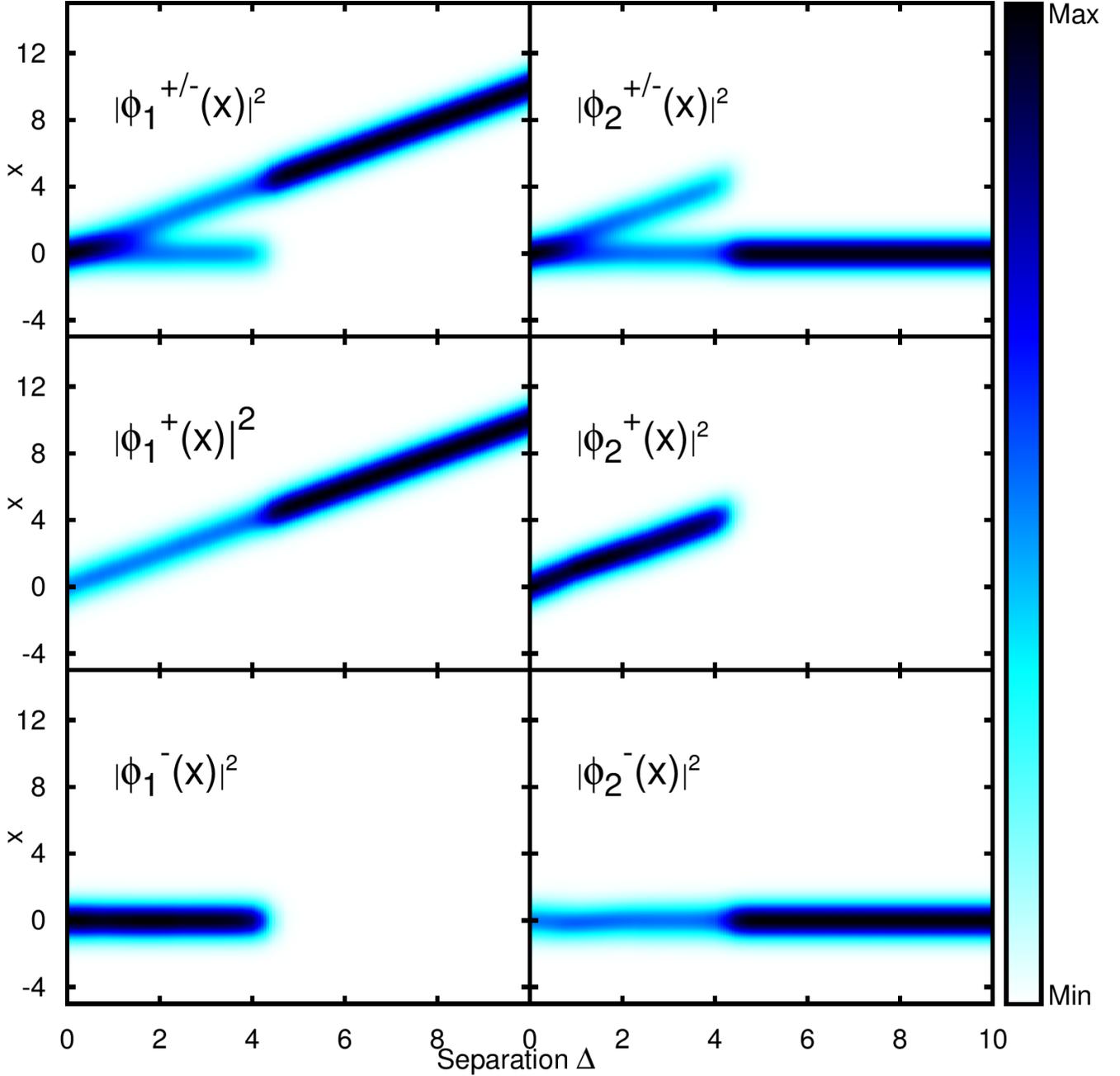}
\caption{Natural orbitals $\vec{\varphi}^{(NO)}_1(x)$ and $\vec{\varphi}^{(NO)}_2(x)$ as a function of the separation $\Delta$. The top panel shows the composite natural orbitals $\phi_k^{+/-,(NO)}(x)=\sum_\alpha \bm{1}^\alpha_x \phi_k^{\alpha,(NO)}(x)$ of the two internal states $\alpha=+$ and $\alpha=-$. The second and third panel depict the component natural orbitals $\phi_k^{\alpha,(NO)}(x)$ of the system in the respective internal state $\alpha=+$ and $\alpha=-$. For the separations, where the fragmentation increases to its maximal value, the composite natural orbital densities are localized in both potential wells (compare top panels and Fig.~\ref{Fig:Densities}, lower panel). For separations $\Delta\gtrsim4.5$, the composite natural orbital densities become localized (top panels), because the component first (second) natural orbital densities in the state $\alpha=-$ ($\alpha=+$) become zero (middle and lower panels). This localization is encompassed by the fragmentation reaching its maximum. All 
quantities 
shown are dimensionless.}
\label{Fig:Orbs}
\end{figure}

\begin{figure}
  \includegraphics[width=\textwidth,angle=-90]{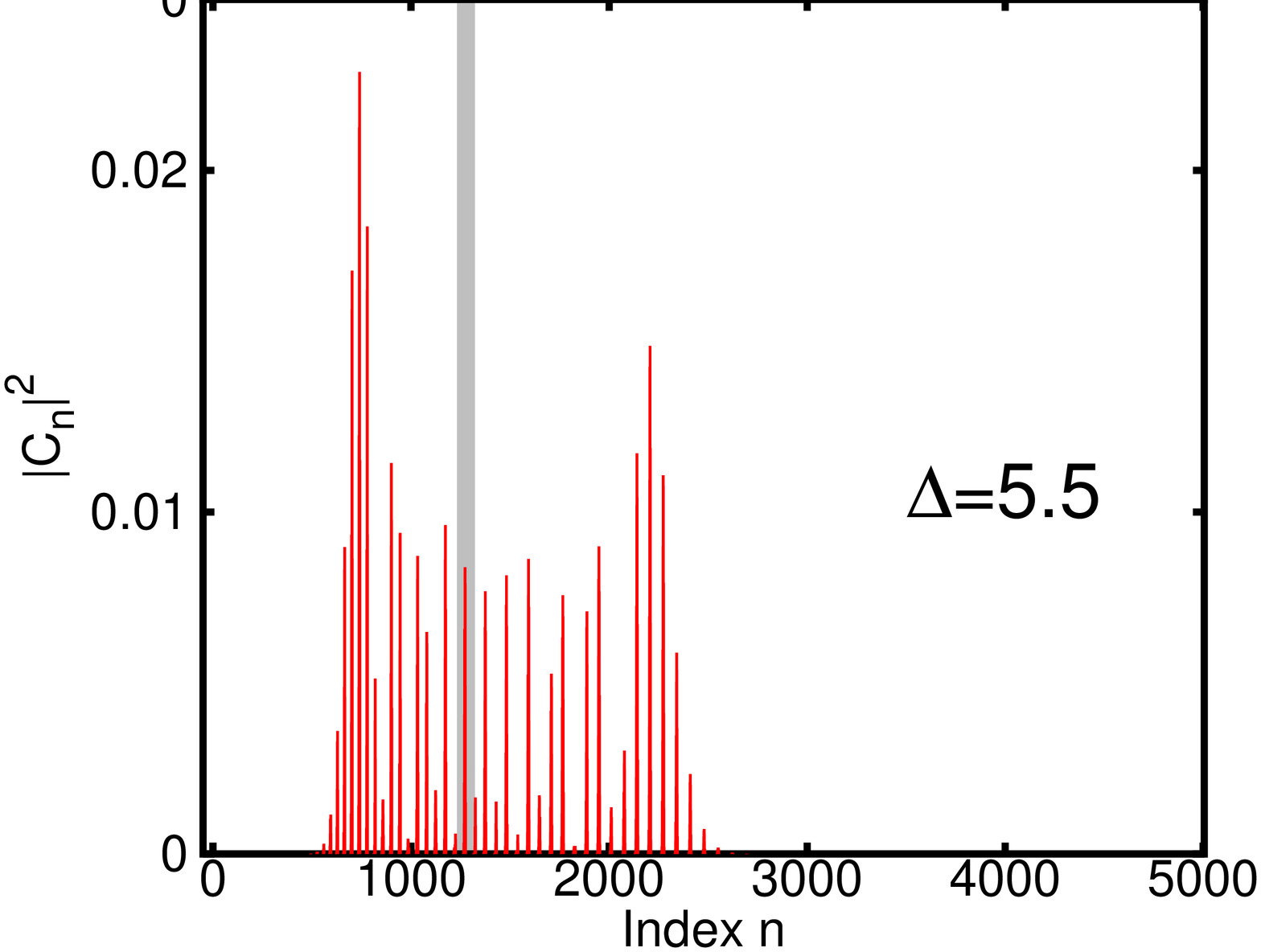}
\caption{Coefficients as a function of the separation $\Delta$. The magnitude of the coefficients $\vert C_n \vert^2$ is plotted (see Ref.~\cite{map} for the formula to compute the index $n$ from the vector $\vec{n}$) for the separations  $\Delta=1,4,5.5$ in the top, middle, and bottom panels, respectively. The thick vertical gray line through all panels shows the configuration $\vert \frac{N}{2}, \frac{N}{2}, 0 \rangle$ for which the bosons are equally distributed in the first two one-particle basis states. For small separations, the system is essentially described by a single coefficient. Encompassing its fragmentation, the distribution of coefficients centers itself around the equally partitioned configuration $\vert \frac{N}{2}, \frac{N}{2}, 0 \rangle$ broadens significantly. All quantities shown are dimensionless.}
\label{Fig:Coefs}
\end{figure}

In this appendix, a detailed analysis of the many-body wavefunction $\Psi$ (Eq.~\eqref{ansatz}) is performed. For this purpose, we plot the natural orbitals $\vec{\varphi}_k^{(NO)}$ which are the eigenfunctions of the reduced one-body density matrix $\bm{\rho}^{(1)}$ in Fig.~\ref{Fig:Orbs} and the coefficients that are the weights of the configurations $\vert \vec{n} \rangle$ contributing to the many-body wavefunction (Eq.~\eqref{ansatz}) in Fig.~\ref{Fig:Coefs}.

\begin{figure}
\vspace*{-1cm}
\includegraphics[angle=-90,width=0.8\textwidth]{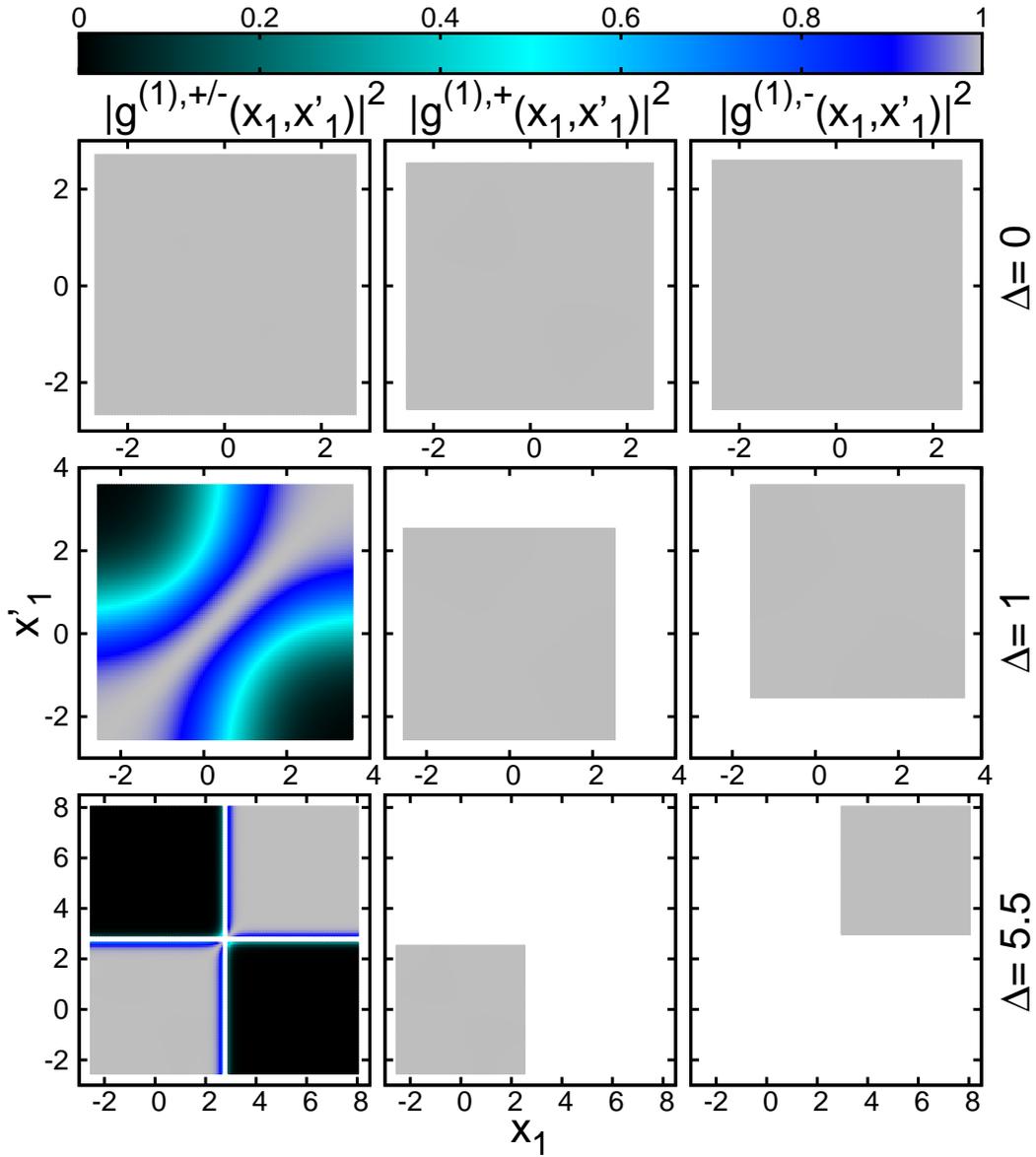}
\caption{Signatures of composite fragmentation in the one-body correlation function as a function of the separation $\Delta$. 
The rows of panels correspond to the separations $\Delta=0$, $\Delta=1$, and $\Delta=5.5$ from top to bottom. The values of fragmentation are, $F=0.004$,$F=0.006$,$F=0.490$, respectively. 
The first column shows the composite correlation function of both internal states $\vert g^{(1),+/-}\vert^2$, the middle column the correlation function of the $\alpha=+$ state, $\vert g^{(1),+}\vert^2$, and the right column the correlation function of the $\alpha=-$ state $\vert g^{(1),-}\vert^2$. The correlations are only plotted for coordinates $(x,x')$ if the component (composite) one-body density at these coordinates is larger than $0.05$, to avoid analyzing component (composite) correlations where there are no particles. While the component correlations exhibit full coherence, i.e., $\vert g^{(1),\alpha}\vert^2 \approx 1$ in the middle and left column, the composite correlation function shows a quick loss of coherence between the components, i.e., $\vert g^{(1),+/-}\vert^2 \approx 0$ on the off-diagonals in the left column: the fragmentation in the system is of ``composite'' type. All quantities shown are dimensionless, see text for further discussion.}
\label{Fig:Corr}
\end{figure}

For small separations before fragmentation sets in, the natural orbitals are ``delocalized'' between both internal degrees of freedom, see Fig.~\ref{Fig:Orbs}. As soon as the separation becomes large enough for the fragmentation to reach is maximal value, the orbitals ``localize'' in one internal state. This sheds further light on the structure of the correlation functions (Fig.~\ref{Fig:Corr}): the components appear to be fully coherent, because each of them is described by an orbital which has practically all its density in a singly component. The composite coherence is lost, because the composite system can only be represented by at least two orbitals (compare Figs.~\ref{Fig:Corr} and \ref{Fig:Orbs}).

From the above analysis of the natural orbitals of the system, one might infer that its fragmentation may be described by single-configurational states, i.e., states which have only a single contributing coefficient $C_{\vec{n}}$ in their many-body wave function $\Psi=\sum_{\vec{n}} C_{\vec{n}}\vert \vec{n};t \rangle$. This, however, is not the case as we shall show now. To this end, we plot the magnitude of the coefficients $\vert C_{\vec{n}} \vert^2$ for various separations $\Delta$ in Fig.~\ref{Fig:Coefs}. For small separations $\Delta$ and absent fragmentation we find an almost perfect single-configurational wavefunction (cf. top panel of Fig.~\ref{Fig:Coefs}). As the separation $\Delta$ increases and fragmentation sets in, the distribution of coefficients gradually broadens while centering itself around the equally partitioned configuration $\vert \frac{N}{2}, \frac{N}{2}\rangle$. As one can clearly see from the middle and lower panel of Fig.~\ref{Fig:Coefs}, the fragmented system is described by many 
configurations and not a single one. Hence, we infer that mean-field theories \cite{PitaSandro} or even multi-orbital mean-field theories \cite{MOMF} are not applicable to the present system.

\clearpage

\begin{thebibliography}{References}
\bibitem{units} To arrive at the dimensionless interaction parameters $\lambda_1=0.0095$, $\lambda_0^-=0.00975$ and $\lambda_0^+ = 0.01$, we first choose a length scale of $L=1\mu m $. The scale of energy for the mass of $^{87}$Rb is $\hbar^2/(m L^2)=116 Hz$ and the scale of time is $mL^2/\hbar=1.37$ milliseconds. According to \cite{Olshanii}, the one-dimensional scattering lengths $\lambda$ are the three-dimensional scattering lengths $a_{3D}$ rescaled by the frequency $\omega_\perp$ of the transversal confinement by $\lambda=2Lm\omega_\perp a_{3D}/\hbar$. We pick the $\alpha=+$ state and map $\lambda_0^+$ to $a_{3D}^{11}=100.4 a_0$, where $a_0$ is the Bohr radius. One obtains $\omega_\perp = \lambda_0 \hbar/(2Lma_{3D}^{11})=687.9$Hz.
\end{thebibliography}
\end{document}